\documentclass[pra, reprint,floatfix,superscriptaddress]{revtex4-2}
\usepackage{amsmath, amssymb, mathtools, bm, bbold, booktabs, pgf} 
\usepackage[export]{adjustbox}
\usepackage{float}
\usepackage[colorlinks=true,citecolor=blue,linkcolor=blue,urlcolor=blue, breaklinks]{hyperref}
\usepackage[mathscr]{eucal}

\begin{document}

\title{Protocols for classically training quantum generative models on probability distributions}

\author{Sachin Kasture}
\affiliation{PASQAL, 7 rue Léonard de Vinci, 91300 Massy, France}

\author{Oleksandr Kyriienko}
\affiliation{Department of Physics and Astronomy, University of Exeter, Stocker Road, Exeter EX4 4QL, United Kingdom}

\author{Vincent E. Elfving}
\affiliation{PASQAL, 7 rue Léonard de Vinci, 91300 Massy, France}

\date{\today}

\begin{abstract}
Quantum Generative Modelling (QGM) relies on preparing quantum states and generating samples from these states as hidden --- or known --- probability distributions. As distributions from some classes of quantum states (circuits) are inherently hard to sample classically, QGM represents an excellent testbed for quantum supremacy experiments. Furthermore, generative tasks are increasingly relevant for industrial machine learning applications, and thus QGM is a strong candidate for demonstrating a practical quantum advantage. However, this requires that quantum circuits are trained to represent industrially relevant distributions, and the corresponding training stage has an extensive training cost for current quantum hardware in practice. In this work, we propose protocols for classical training of QGMs based on circuits of the specific type that admit an efficient gradient computation, while remaining hard to sample. In particular, we consider Instantaneous Quantum Polynomial (IQP) circuits and their extensions. Showing their classical simulability in terms of the time complexity, sparsity and anti-concentration properties, we develop a classically tractable way of simulating their output probability distributions, allowing classical training to a target probability distribution. The corresponding quantum sampling from IQPs can be performed efficiently, unlike when using classical sampling. We numerically demonstrate the end-to-end training of IQP circuits using probability distributions for up to 30 qubits on a regular desktop computer. When applied to industrially relevant distributions this combination of classical training with quantum sampling  represents an avenue for reaching advantage in the noisy intermediate-scale quantum (NISQ) era. 
\end{abstract}\maketitle

\maketitle

%======   INTRO =========

\section{Introduction}

Recent breakthrough works in quantum computing demonstrated an improved scaling for sampling problems, leading to so-called the quantum supremacy \cite{Arute2019}. While an exact boundary of classical simulation is still to be established \cite{Pan2021}, for carefully selected task of random circuits sampling and boson sampling \cite{Zhong2020,Madsen2022} one achieves an exponential separation for the task of generating samples (bit strings from measurement read-out). Generally, the task of generating samples from underlying probability distributions is the basis of generative modelling, and represents a highly important part of classical machine learning. Its quantum version --- Quantum Generative Modelling (QGM) --- relies on training quantum circuits as adjustable probability distributions that can model sampling from some \textit{particular} desired distribution \cite{Georgescu2022,Kumar2022,Dallaire-Demers2018,Hu2019,Benedetti2019,Verdon2019}.
%%%
\begin{figure}[h!]
    \centering 
    \includegraphics[width=1.0\linewidth]{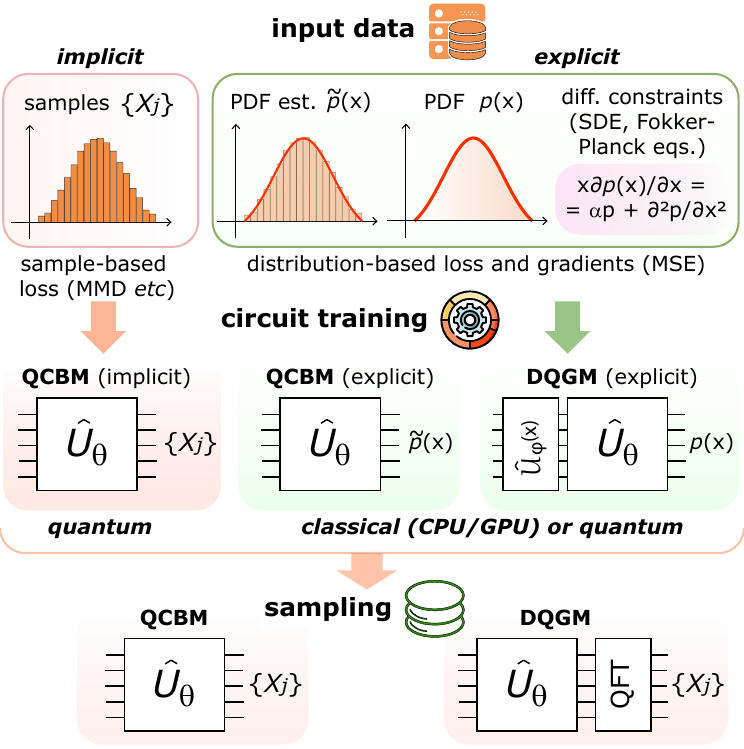}
    \caption{Schematic of different steps in training a quantum generative model using QCBM and DQGM architectures. In a conventional QCBM setup, the loss and gradient estimation is done using input data samples directly, while in our work we focus on an explicit version of QCBM and DQGM, allowing for classical training.
\label{fig_idea_summary}}
\end{figure}
%%%
QGM exploits the inherent superposition properties of a quantum state generated from a parameterized unitary, along with the probabilistic nature of quantum measurements, to efficiently sample from a trainable model. QGM have potential applications in generating samples from solutions of stochastic differential equations (SDEs) for simulating financial and diffusive physical processes \cite{Paine2021,kubo_variational_2021}, scrambling data using a quantum embedding for anonymization \cite{landsman_verified_2019,li_quantum_2019,harris_benchmarking_2022}, generating solutions to graph-based problems like maximum independent set or maximum clique \cite{banchi_molecular_2020,ebadi_quantum_2022}, among many others. Given the successes of quantum sampling this makes QGM a promising contender for achieving a quantum advantage. However, to date demonstrating QGM of practical significance has eluded the field as training specific generative models is a complex time-intensive task.

The key asset of quantum generative modelling is that quantum measurements (collapse to one of eigenstates of a measurement operator) provide a new sample with each shot. Depending on the hardware platform, generating one sample can take on the order of a few hundreds of microseconds to several milliseconds \cite{Gross2017,Graham2022,Henriet2020,Dicarlo2009,Harrigan2021,Martinez2016,Figgatt2017}. For large probability distributions, represented by entangled 50+ qubit registers, performing the classical inversion is indeed much more costly. However, often the challenge comes from the inference side, when quantum circuits are required to match specific distributions. Typically, training of quantum generative models utilizes gradient-based parametric learning, similarly to training of deep neural networks \cite{Rumelhart1986}.  Parameterized quantum circuits (also referred as quantum neural networks --- QNNs) are trained by estimating the gradient of gate parameters $\bm{\theta}$. Moreover, the gradient of a full QGM loss has to be estimated with respect to $\bm{\theta}$. For quantum devices this can be done by the parameter-shift rule and its generalization \cite{Mitarai2018,Schuld2019,kyriienko_generalized_2021}, where number of circuit estimation increases linearly with the number of parameters. The overall training cost corresponds to the measurement of the loss function at each iteration step. In the case of quantum circuit Born machine the loss may correspond to Kullback-Leibler(KL) divergence \cite{Benedetti2019}, Sinkhorn divergence \cite{Coyle2020} or maximum mean discrepancy (MMD) \cite{Liu2018}, and may require extensive sampling for resolving the loss as an average. For quantum generative adversarial networks (QGAN) the loss minimization is substituted by the minimax game \cite{Zoufal2019,Huang2021} requiring multi-circuit estimation. In all cases the convergence is not guaranteed due to exponential reduction of gradients \cite{McClean2018BarrenLandscapes}. 
%%%
\begin{figure*}[th!]\centering
\includegraphics[width=1.0\linewidth]{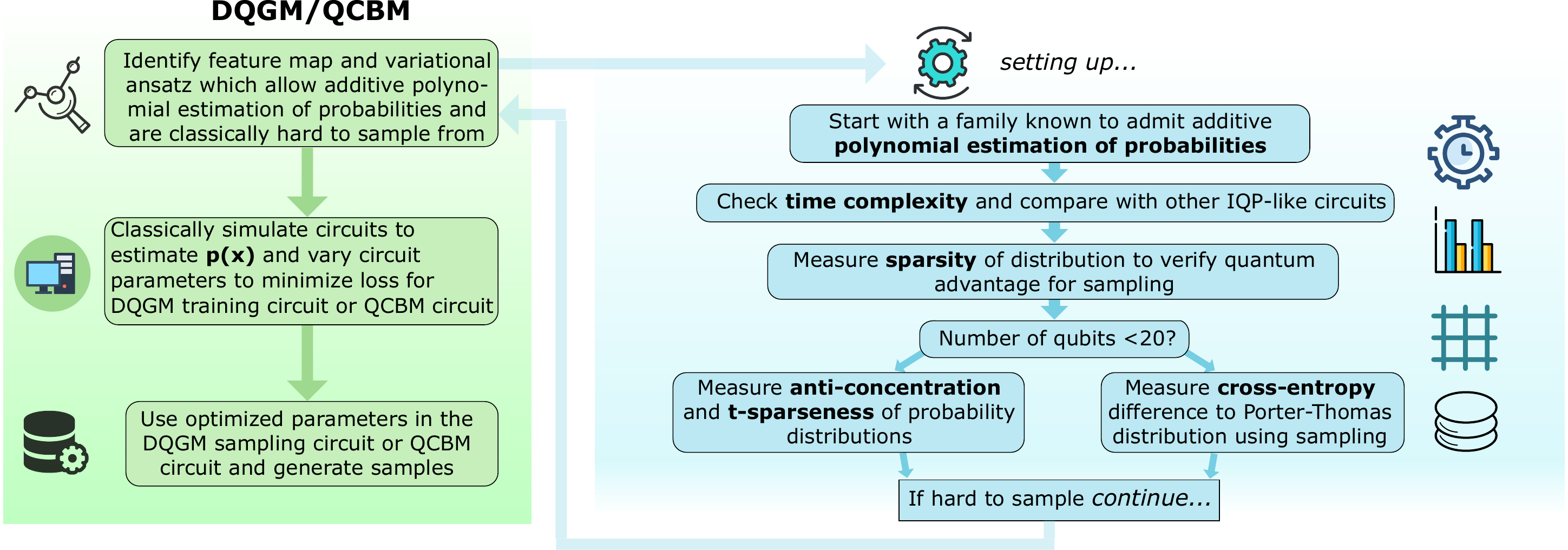}
\caption{Workflow used for classical training of quantum samplers, both in the differentiable quantum generative modelling (DQGM) and the quantum circuit Born machine (QCBM) setting. First, we identify circuits $\hat{\mathcal{U}}$ suitable for classically tractable probability estimation and hard sampling (see the right chart). For this we: 1) check that a circuit admits additive polynomial estimation of probability; 2) compare the time-complexity for exact probability calculation with known circuit families; 3) verify the sparseness of the probability distributions, depending on the number of qubits $n$; 4) measure anti-concentration/t-sparseness or cross entropy difference using sampling. Analysing these properties we confirm if $\hat{\mathcal{U}}$ admits classical training and computationally hard sampling. After the classical training is performed variationally by minimizing DQGM/QCBM loss, we use optimized parameters for quantum sampling circuits.}
\label{training_strategy}
\end{figure*}
%%%

In this work we investigate the possibility of training the parameters of quantum generative models classically, while still retaining the quantum advantage in sampling \cite{patent}. For instance, the ability of classical training for a different paradigm was shown for Gaussian Boson Sampling (GBS) devices \cite{Banchi2020a}, but under certain conditions of fixing an initial set of samples and non-universal operation. For the digital quantum computing operation, results from previous works \cite{Pashayan2020,VandenNest2011} motivate the possibility that estimating probability density classically can be feasible without losing the sampling complexity. We explore this possibility in more detail and use this further to develop methods to train circuits classically to output a desired distribution using a gradient-based approach. We show that our method is feasible using numerics for up to 30 qubits on a regular desktop computer. We explore different families of quantum circuits in detail and perform numerical studies to study their sampling complexity and expressivity. For expressivity studies in particular, we look at training a Differentiable Quantum Generative Model (DQGM) \cite{Kyriienko2022,kyriienko_solving_2021,Paine_quantum_2022} architecture which allows training in the latent (or `frequency') space, and sampling in the bit-basis. This presents a good testing ground for applying the proposed method to explicit quantum generative models. We also show that QCBMs can be trained classically for certain distributions, while still hard to sample. Our protocols contribute towards tools for achieving the practical advantage in sampling once the target distributions are chosen carefully. We highlight the differences between well known strategies for QGM and the method discussed in the paper in Fig.~\ref{fig_idea_summary}.
%%%
%\begin{figure}[t]\centering
%\includegraphics[width=1.0\linewidth]{Fig4.pdf}
%\caption{Figure shows the basic strategy we use for classical training of quantum samplers, both in the DQGM and the QCBM setting.}
%\label{training_strategy}
%\end{figure}
%%%

\section{Methods}

\subsection{Preliminaries: QCBM and DQGM as implicit vs explicit generative models}

Generally, there are two types of generative models. \textit{Explicit} generative models assume a direct access (or `inference') to probability density functions (PDF). At the same time, \textit{implicit} generative models are described by hidden parametric distributions, where samples are produced by transforming randomness via inversion procedure. These two types of models have crucial differences. For example, training explicit models involves loss functions measuring distances between a given PDF, $p_{\mathrm{target}}(x)$ and a model PDF, $p_{\mathrm{model}}(x)$, for example with a mean square error (MSE) loss which is defined as
\begin{equation}
    \mathcal{L}_{\mathrm{MSE}} = \sum_x|p_\mathrm{model}(x )- p_\mathrm{target}(x)|^2
\end{equation}
where explicit knowledge of the $p_{\mathrm{target}}(x)$ is used.
On the other hand, training implicit models involves comparing the samples generated by the model with given data samples (e.g. with a  MMD loss  \cite{Liu2018}). The MMD loss is defined as
\begin{equation}
\begin{split}
\mathcal{L}_{\mathrm{MMD}}&= \underset{x\sim p_{\mathrm{model}},y\sim p_{\mathrm{model}} }{\mathbb{E}} K(x,y)\\
&-2\underset{x\sim p_{\mathrm{model}},y\sim p_{\mathrm{target}} }{\mathbb{E}} K(x,y)\\
&+ \underset{x\sim p_{\mathrm{target}},y\sim p_{\mathrm{target}} }{\mathbb{E}} K(x,y)\\
\end{split}
\label{eq_MMD}
\end{equation}
where $K(x,y)$ is an appropriate kernel function. The MMD loss measures the distance between two probability distributions using samples drawn from the respective distributions as shown in the above equation. 
%Typically generative modelling involves minimizing a loss function which measures the `distance' between samples from a target probability distribution ($p_{target}(x)$) and samples coming from a model probability distribution ($p_{model}(x)$), where samples $\{X_j\}$ in case of qubits consists of bit-strings of 0's and 1's. Various measures of distance like KL-divergence or MMD can be used. Therefore training the model requires estimating the value of $p_{model}(x)$ or sampling from this distribution. 
In the context of QGM, QCBM is an excellent example of implicit training where typically a MMD like loss-function is used. On the other hand, recent work showcases how \textit{explicit} quantum models such as DQGM \cite{Kyriienko2022} and Quantum Quantile Mechanics \cite{Paine2021}  benefit from a \textit{functional} access to the model probability distributions, allowing input-differentiable quantum models  \cite{kyriienko_solving_2021, Paine_quantum_2022} to solve stochastic differential equations or to model distributions with differential constraints. 

Let us consider a quantum state $|\Psi\rangle$ created by applying a quantum circuit $\hat{\mathcal{U}}$ (which can be parameterized) to a zero basis state. For a general $\hat{\mathcal{U}}$, simulating the output PDF values that follow the Born rule $p_{\mathrm{model}}(x)=|\langle x|\Psi\rangle|^2$ \emph{and} producing samples from $|\Psi\rangle$ are both classically hard. But what if \textit{estimating} the PDF for certain $\hat{\mathcal{U}}$ to sufficient accuracy is classically tractable? In this case one can use an explicit training, and at the inference stage have access not only to probabilities but also the capacity to sample efficiently via quantum measurements. This scenario describes a potential for classical training of quantum generative models. 

To enable the classical training, we propose a strategy described in a schematic shown in Fig.~\ref{training_strategy}. Here, our goal is finding circuits satisfying 'classical training + quantum sampling' conditions.

\subsection{$\pi$-simulable circuits that are hard to sample}

We note that certain families of quantum circuits, including Clifford \cite{Gottesman1998TheComputers,VanDenNest2010} and match-gate sequences \cite{Jozsa2008,Terhal2001,Brod2016}, admit a classical tractable \textit{estimation} of probabilities $p_{\mathrm{model}}(x)$. At the same time, for these circuits the generation of samples is also classically `easy', thus limiting the potential for achieving a quantum advantage. However, there exist families of quantum circuits that allow for complexity separation between the two tasks. For instance, in Ref.~\cite{Pashayan2020} the authors show that one can estimate probabilities for IQP (Instantaneous Quantum Polynomial) circuits \cite{Shepherd2010,Shepherd2010_2} up to an additive polynomial error, while retaining a classical hardness for sampling. For a typical IQP circuit with input $|0^{\otimes n}\rangle$, the amplitude to obtain a certain bit-string $x$ at output is given by
\begin{equation}
    \Psi(x)=\langle x |\hat{H}\hat{U}\hat{H} |0^{\otimes n}\rangle
    \label{eq_IQP}
\end{equation}
where $\hat{U}$ consists of single and 2-qubit Z-basis rotations and $\hat{H}$ represents the Hadamard gate applied to all the qubits. Moreover, it is known to be classically hard to even \textit{approximately} sample from IQP circuits in an average case \cite{Bremner2016}. Therefore, such circuits offer an opportunity for explicit training of a quantum generative model, and a potential for quantum advantage in sampling.
%%%
%\begin{figure}[t]\centering
%    \centering
 %   \includegraphics[width=0.95\columnwidth]{Fig3.pdf}
%    \caption{General strategy proposed in this work to study properties quantum circuits $\hat{\mathcal{U}}$ for generative modelling at increasing number of qubits $n$. Once a family $\hat{\mathcal{U}}$ that admits additive polynomial estimation of probability is chosen, we can compare first the time-complexity for exact probability calculation with known circuit families. To verify the sparseness of the probability distributions, depending on the number of qubits, we measure anti-concentration/t-sparseness or cross entropy difference using sampling. Analysing these properties we confirm if $\hat{\mathcal{U}}$ admits classical training and computationally hard sampling.}
%    \label{strategy}
%\end{figure}
%%%
IQP circuits by its structure are also strongly related to the so-called forrelation problem \cite{Aaronson2015,Bansal2021,Havlicek2019}, they only differ by an additional $\hat{H}$ layer in the middle. This corresponds to calculating an overlap between states defined as
\begin{equation}
    \Phi = \langle 0^{\otimes n}|\hat{H}\hat{U}_2 \hat{H} \hat{U}_1 \hat{H}|0^{\otimes n}\rangle
    \label{Eq_forrelation}
\end{equation}
after the action of quantum circuit $\hat{U}_{\mathrm{F}}:=\hat{H}\hat{U}_2 \hat{H} \hat{U}_1 \hat{H}$, where $\hat{U}_1,\hat{U}_2$ are circuits that consist of Z-basis rotations $\hat{R}_z$ and Ising-type propagators $\hat{R}_{zz}$. Hereafter, $\hat{H}$ is a layer of Hadamard gates applied to all $n$ qubits. It was shown that $\Phi$ can be calculated efficiently classically. Therefore, from the squared forrelation $|\Phi|^2$ one can estimate the probability to obtain $|0^{\otimes n}\rangle$ at the output after the action of $\hat{U}_{\mathrm{F}}$, provided the input is $|0^{\otimes n}\rangle$. At the same time, in the following we show that the variant of forrelation can be used for achieving the sampling advantage.

Classical estimation of probability using the forrelation has been shown to be possible \cite{Bravyi2021} under the condition that the circuit's entangling properties are constrained. To understand this, we use the concept of connectivity in graph theory. Consider a graph $G$ with $n$ nodes, where each node represents a qubit in a quantum circuit. Two nodes are connected by an edge if there is a 2-qubit entangling gate between the corresponding qubits in the circuit. Typically IQP circuits have all-to-all connectivity, usually by using a two-qubit entangling gate. However, if we restrict the connectivity such that the resulting connectivity graph is bipartite, we can obtain probabilities up to an additive polynomial error classically efficiently for these `extended-IQP' circuits. More concretely, whenever the connectivity graph can be partitioned into 2 disjoint subsets such that the tree-decomposition of each of the subsets has a small tree-width, then a classical algorithm is possible with a runtime of $O(n4^w\epsilon^{-2})$ where $w$ is the maximum tree-width of the decomposition  \cite{Bravyi2021}, $n$ is the number of qubits and $\epsilon$ is error in the estimated probability. 
We show examples of a bipartite graph with 4 nodes and a complete graph in Appendix A, as well as the corresponding circuits. In the rest of the text, we use the term extended-IQP circuits to mean these quantum circuits which have a bipartite connectivity graph and an additional Hadamard layer between the set of commuting gates.
%%%
%\begin{figure}[!htbp]
%    \centering
%    \includegraphics[width=\linewidth]{Fig1.pdf}
%    \caption{Schematic diagram showing how the Quantum Circuit Born Machine (QCBM) and Differentiable Quantum Generative Model (DQGM) can be trained either on a quantum computer or, for certain circuit unitaries, on a classical computer, where quantum sampling is performed either way.\label{Fig1}}
%\end{figure}
%%%
%%%
\begin{figure}[t]
\centering
    \includegraphics[width=1.0\linewidth]{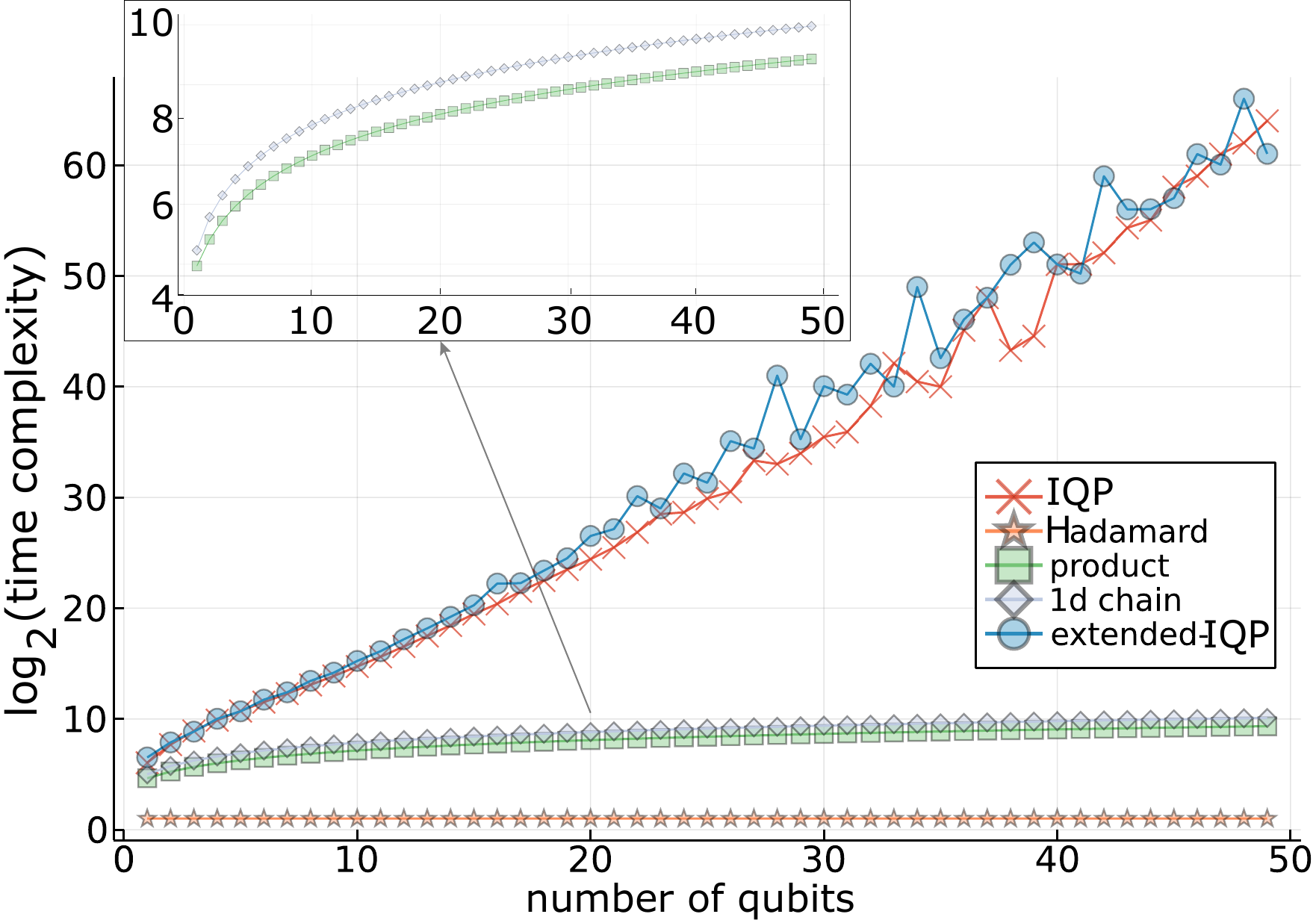}
\caption{Time-complexity for different families of circuits shown on a logarithmic scale. We observe that for IQP and extended-IQP the time complexity scales exponentially. Simple Hadamard circuits has a constant complexity. For product circuits and IQP circuits in a 1D chain, we see that the complexity has a saturation behavior and thus only a polynomial increase (also see inset).}
\label{time_complexity}
\end{figure}
%%%

%-------  DQGM training -------------

\subsection{Training a DQGM efficiently classically}

We focus on DQGM circuits as an explicitly-trained generative methodology, although the ideas hold also for architectures typically considered as implicit, like QCBM \cite{Liu2018,Coyle2020} or QGAN \cite{Lloyd2018}. DQGMs allow for separation of training and the sampling stages and allow leveraging of frequency taming techniques like feature map sparsification, qubit-wise learning and Fourier initialization for improving and simplifying the training. DQGM also naturally allows for generative modelling for sampling from solutions to stochastic differential equations, inspired by Physics-inspired Neural Networks (PINN) \cite{Raissi2019Physics-informedEquations,Karniadakis2021Physics-informedLearning} and derivative quantum circuit (DQC)-like \cite{Kyriienko2022} approaches for finding solutions to the time-dependent probability density function and sampling from that. 
The training part of the DQGM consists of a kernel $\hat{U}_{\phi}(x)$ followed by a variational circuit $\hat{U}_{\theta}$. Following \cite{Kyriienko2022}, we define $\hat{U}_{\phi}(x)$ as
\begin{equation}
    \hat{U}_{\phi}(x) = \prod_{i=1}^n \left[\hat{R}_i^z\left(\frac{2\pi x}{2^i}\right) \right] \hat{H}
    \label{Eq_fm}
\end{equation}
where $\hat{R}_i^z$ are single qubit rotation gates around Z axis, which are preceded by $\hat{H} = \prod_{i=1}^n \hat{H}_i$ as the layer of single qubit Hadamard gates. The operator $\hat{U}_{\phi}(x)$ maps an initial state $|\phi\rangle$ (taken as a state of $|0^{\otimes n}\rangle$ for all qubits) to a product state $|\Tilde{x}\rangle$, which is a latent space representation of the variable $x$. The transform $\hat{U}_{T\phi}$ transforms this to a binary state $|x\rangle$ as a bijection. This circuit is dependent on the feature map and for the above described map (Eq.~\ref{Eq_fm}) corresponds to the inverse quantum Fourier transform (iQFT). The training stages can be described as a sequence of steps
\begin{equation}
    |\phi\rangle \xrightarrow[\text{}]{\hat{U}_{\phi}(x)}|\Tilde{x}\rangle\xrightarrow[\text{}]{\hat{U}_{T\phi}}|x\rangle \xrightarrow[\text{}]{\hat{U}_\theta \hat{U}_{T\phi}^{\dagger}}P_{\mathrm{train}}(0^{\otimes n}) ,
    \label{Eq_dqgm_train}
\end{equation}
where $P_{\mathrm{train}}(0^{\otimes n})$ denotes finding classically the probability of observing $0^{\otimes n}$ bitstring. 
%%%

%%%
Similarly for the sampling stage we have
\begin{equation}
    |\phi\rangle \xrightarrow[\text{}]{\hat{U}_{T\phi}\hat{U}_{\theta}^\dagger}P_{\mathrm{sampling}}(|x\rangle \langle x|),
    \label{Eq_dqgm_sampling}
\end{equation}
where at the end we perform quantum measurements in the computational basis. It can be shown that for a given $x$, $P_{\mathrm{train}}(0^{\otimes n})=P_{\mathrm{sampling}}(|x\rangle \langle x|)$. We therefore train $\hat{U}_{\theta}$ so that $P_{\mathrm{train}}(0^{\otimes n})=p_{\mathrm{target}}(x)$ for all values of $x$, where $p_{\mathrm{target}}(x)$ is the target probability distribution we want to sample from.

Next, to be able to train the DQGM efficiently, we now rewrite the corresponding circuits $\hat{U}_\phi(x)\hat{U}_\theta$ as an extended-IQP circuit. The extended-IQP has the form $\hat{H}\hat{U}_1\hat{H}\hat{U}_2\hat{H}$ where the depth is doubled. Thus we can assign $\hat{U}_\phi(x)=\hat{H}\hat{U}_1(x)$ and $\hat{U}_\theta=\hat{U}_1\hat{H}\hat{U}_2\hat{H}$. Here we have split up $\hat{U}_1$ so that part of it can be used as a feature map (consisting of single qubit $x$-dependent $\hat{R}_z$ rotations), and the remaining consisting of $\hat{R}_z$ and $\hat{R}_{zz}$ with bi-partite connectivity as a part of the variational ansatz $\hat{U}_{\theta}$.  Similarly DQGM training circuit can be written as an IQP circuit by setting $\hat{U}_{\phi}(x)=\hat{H}\hat{U}_1$ and $\hat{U}_\theta=\hat{U}_2 \hat{H}$. Note that in case of IQP $\hat{U}_1$ and $\hat{U}_2$ do not have to be limited to bipartite connectivity. 

We now show that gradients with respect to circuit parameters $\theta$ can also be classically estimated efficiently for an extended-IQP circuit. Here we assume a DQGM setting, although the conclusions hold for a QCBM as well (which we show in the next section). We have seen that for an extended-IQP circuit, we can estimate the probability of having output $|0^{\otimes n}\rangle$, i.e $p_{\mathrm{model}}(x)=p(0)=\mathrm{tr}\{|0^{\otimes n}\rangle \langle 0^{\otimes n}|\hat{\rho}_{\mathrm{out}} \}$ can be estimated efficiently, where $\hat{\rho}_{\mathrm{out}}$ is the output density matrix. Suppose we have a gate $\hat{U}_{2,j}=e^{i\theta_j \hat{P}_j/2}$ in $\hat{U}_2$ where $\hat{P}_j$ are the Pauli operators. Then the gradients using parameter-shift rule \cite{Mitarai2018} can be written as
\begin{equation}
    \begin{split}
        &\frac{\partial p_{\mathrm{model}}(x)}{\partial \theta_j}=\mathrm{tr}[|0^{\otimes n}\rangle \langle 0^{\otimes n}|\hat{H}\hat{U}_{2,l:j+1}\hat{U}_{2,j}(\pi/2)\hat{\rho}_j\\
        &\hat{U}^\dagger _{2,j}(\pi/2)\hat{U}^\dagger_{2,l:j+1}\hat{H}]-\mathrm{tr}[|0^{\otimes n}\rangle \langle 0^{\otimes n}|\hat{H} \hat{U}_{2,l:j+1}\\
        &\hat{U}_{2,j}(-\pi/2)\hat{\rho}_j
        \hat{U}^\dagger _{2,j}(-\pi/2)\hat{U}^\dagger_{2,l:j+1}\hat{H}],\\
    \end{split}
    \label{Eq_dpdtheta}
\end{equation}
where $\hat{\rho}_j=\hat{U}_{2,1:j}\hat{H}\hat{U}_1\hat{H}|0^{\otimes n}\rangle \langle 0^{\otimes n}|\hat{H}\hat{U}_1^\dagger \hat{H}\hat{U}_{2,1:j}^{\dagger}$. Since both the terms in the above equation are probabilities of obtaining $|0^{\otimes n}\rangle$ from the extended-IQP circuit, they can be estimated efficiently. A similar approach also works for a IQP circuit.
%%%

%%%
Note that in the context of solving SDEs \cite{Paine2021,Kyriienko2022,Alghassi2022}, it can be shown that differentials with respect to $x$ like $dp_{\mathrm{model}}(x)/dx$ and higher order derivatives can also be estimated efficiently classically.

We now show how gradients of probabilities with respect to circuit parameters $\theta$ can be estimated classically efficiently. Working in the DQGM setting and using the extended-IQP architecture, we know that 
\begin{equation}
    p(x)=P(0)=|\Phi|^2 ,
    \label{Eq_px}
\end{equation}
where we use the definition of forrelation $\Phi$ from Eq.~\eqref{Eq_forrelation} and write $\hat{U}_1=\hat{U}_1(\theta_1)$ and $\hat{U}_2=\hat{U}_2(\theta_2)$ as parameterized trainable layers. Following Ref.~\cite{Bravyi2021}, $\Phi$ can be written as
\begin{equation}
    \Phi = \sum_y P(y)R(y)
    \label{Eq_MC}
\end{equation}
where $R(y)=\frac{\langle \beta| y \rangle}{\langle \alpha | y\rangle}$ and $|\alpha\rangle = (\hat{H}_A\otimes \hat{I}_B)\hat{U}_1\hat{H}|0^{\otimes n}\rangle$ and $|\beta\rangle = (\hat{H}_B\otimes \hat{I}_A)\hat{U}_2^\dagger \hat{H}|0^{\otimes n}\rangle$ and $P(y)=|\langle y | \alpha\rangle|^2$. Thus $\Phi$ can be estimated by sampling from $P(y)$. One way of calculating the gradients with respect to $\theta$ would be to use the parameter shift rule, which has been described in Eq.~\eqref{Eq_dpdtheta}. However this would involve re-sampling from $P(y)$ with shifted circuit parameters. For large number of parameters this becomes inefficient. Instead we now use the following approach: The derivative of $|\Phi|^2$ can be written as
\begin{equation}
    \frac{d|\Phi|^2}{d\theta}=\Phi^*\frac{d\Phi}{d\theta} + \frac{d\Phi^*}{d\theta}\Phi .
    \label{Eq_dforr2dtheta}
\end{equation}
%
%%%

%%%
\begin{figure}[t]\centering
\includegraphics[width=1.0\linewidth]{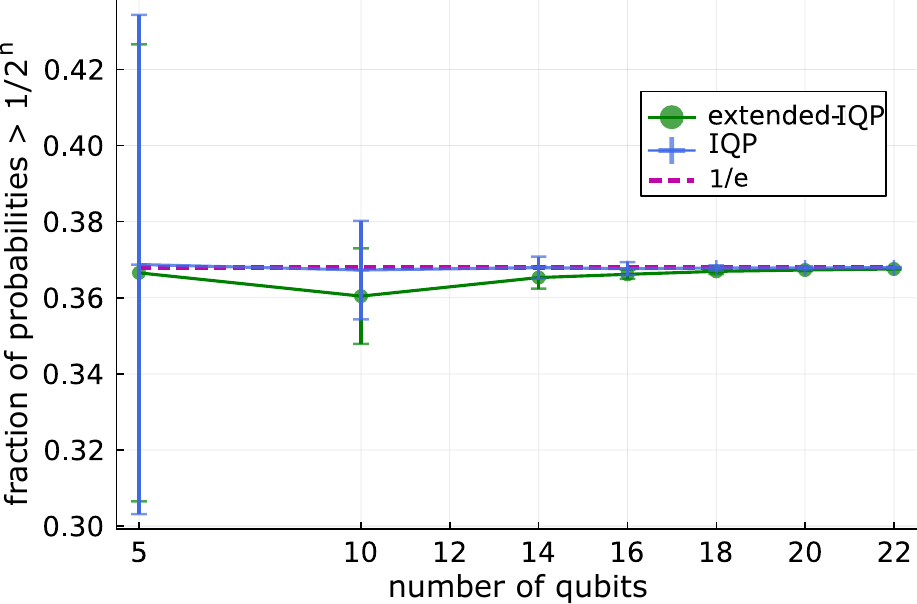}
\caption{Figure shows the anti-concentration properties of extended-IQP which has a bipartite connectivity graph and IQP with full connectivity. The fraction of probabilities $\ge 1/2^n$ is very close to $1/e$ which is shown by dotted line.The error bars show the standard deviation over 100 instances of unitary matrices for each family.}
\label{anticoncentration}
\end{figure}
Using Eq.~\eqref{Eq_MC} we differentiate $\Phi $ with respect to $\theta$ and get
\begin{equation}
    \frac{d\Phi}{d\theta} = \sum_y \frac{dP(y)}{d\theta}R(y) + \sum_y P(y)\frac{dR(y)}{d\theta}.
    \label{Eq_dforrdtheta}
\end{equation}
The second term can be estimated using the same samples used to estimate $\Phi$. For the first term we can write 
\begin{equation}
    \sum_y\frac{dP(y)}{d\theta}R(y) = \sum_y P(y)\frac{1}{P(y)}\frac{dP(y)}{d\theta}R(y).
    \label{Eq_diffMC}
\end{equation}
Now using samples drawn from $P(y)$ we can estimate the value of $\frac{1}{P(y)}\frac{dP(y)}{d\theta}R(y)$. Thus this method avoids the need for repeated re-sampling from $P(y)$ to estimate gradients. Eqs.~\eqref{Eq_dforr2dtheta}-\eqref{Eq_diffMC} can be then used to estimate the gradients for $|\Phi|^2$.

Although we have focused on using estimating probability densities classically for training, for solving general QGM problems we would also like to be be able to estimate more general observables or `cost functions'. These can involve different sets of operators other than the zero state overlap. We now show that for an extended-IQP circuit, also more general expectation values can be calculated classically efficiently. For example, let us consider an expectation value of the operator $\Gamma=\sum_{i,j} Z_i Z_j$, where $i,j$ index the qubits. These terms occur in an Ising Hamiltonian which is used in the formulation of binary optimization problems using digital or analog quantum devices. For the extended-IQP circuit we can write
\begin{equation}
    \langle \Gamma \rangle = \sum_{i,j}\langle 0^{\otimes n}|\hat{H}\hat{U}_p^\dagger \hat{H}\hat{U}_q^\dagger \hat{H}Z_iZ_j\hat{H}\hat{U}_q\hat{H}\hat{U}_p\hat{H}|0^{\otimes n}\rangle .
    \label{Eq_gamma}
\end{equation}
\begin{figure}[t]\centering
\includegraphics[width=1.0\linewidth]{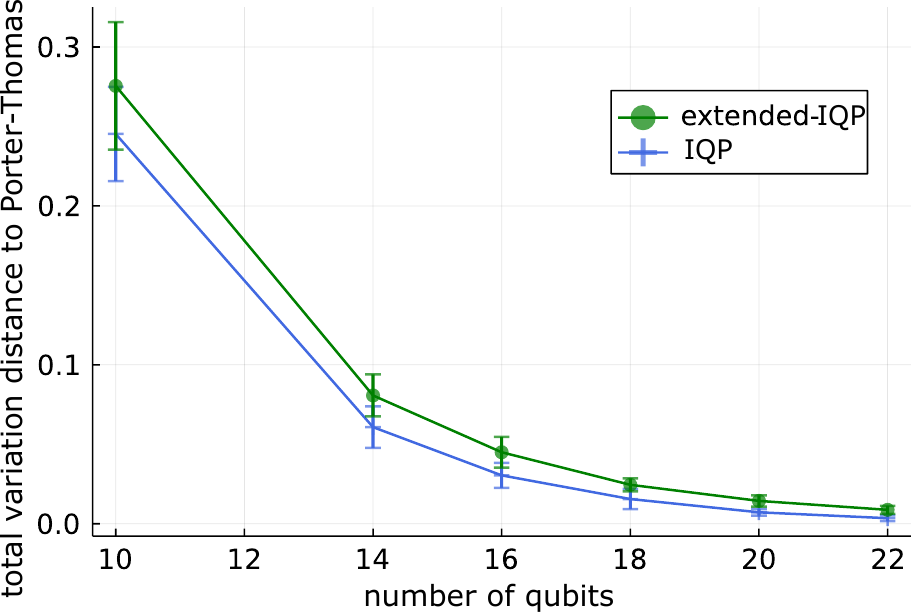}
\caption{Figure shows the total variation distance with respect to the Porter-Thomas distribution. The distance rapidly approaches zero as we increase the number of qubits. The plot shows the mean and the variance for the total variation distance over 100 distributions for each qubit number. }
\label{Porter_Thomas}
\end{figure}
We now consider a single term in the summation and consider the term $Z_1 Z_2$ and writing $\hat{I}_{12}=\sum_{a,b\in\{0,1\}}|z_az_b\rangle\langle z_az_b|$, we get
\begin{equation}
\begin{split}
    \langle Z_1 Z_2 \rangle &= \sum_{a,b,c,d \in \{0,1\}} \langle 0^{\otimes n}|\hat{H}\hat{U}_p^\dagger \hat{H}|z_az_b\rangle\\
    &\langle z_a z_b|\hat{U}_q^\dagger \hat{H}Z_1Z_2\hat{H}\hat{U}_q|z_cz_d
    \rangle \langle z_cz_d|\hat{H}\hat{U}_p\hat{H}|0^{\otimes n}\rangle
\end{split}
\label{Eq_meanz1z2}
\end{equation}
A single term in the summation can be written as
\begin{equation}
\begin{split}
    \langle \hat{Z_1}\hat{Z_2}\rangle_{abcd}&=\langle z_a z_b|\hat{U}_{12}|z_cz_d\rangle \langle 0^{\otimes n}|\hat{H}\hat{U}_p^\dagger \hat{H}|z_az_b\rangle\\
    &\langle z_cz_d |\hat{H}\hat{U}_p\hat{H}|0^{\otimes n}\rangle
    \label{Eq_sum1}
    \end{split}
\end{equation}
where $\hat{U}_{12}=\hat{U}^\dagger _{q,12}\hat{H}_1\hat{H}_2Z_1Z_2\hat{H}_1\hat{H}_2\hat{U}_{q,12}$ with $\hat{H}_i$ being the Hadamard gate acting on qubit $i$ and the terms in $\hat{H},\hat{U}_q$ which do not contain terms for qubits $1,2$ commute through $Z_1,Z_2$ and meet their conjugates and are converted to identity. This term can be calculated classically efficiently since it is only a 2-qubit overlap integral. The second term in the above product can be written as
\begin{equation}
\begin{split}
    &\langle 0^{\otimes n}|\hat{H}\hat{U}_p^\dagger \hat{H}|z_az_b\rangle\langle z_cz_d|\hat{H}\hat{U}_p\hat{H}|0^{\otimes n}\rangle =\\
    &=\langle 0^{\otimes n}|\hat{H}\hat{U}_p^\dagger \hat{\delta}\hat{U}_p\hat{H}|0^{\otimes n}\rangle ,
\label{Eq_sum2}
\end{split}
\end{equation}
where $\hat{\delta}=\hat{H}|z_az_b\rangle \langle z_c z_d|\hat{H}$ which is a tensor product operator. The authors of Ref. \cite{Bravyi2021} prove that this term can be also calculated classically efficiently up to an additive polynomial error. Hence $\langle Z_1 Z_2 \rangle$ and consequently $\langle \Gamma \rangle$ can be calculated classically efficiently. It can be similarly shown that expectation values for operators like $\sum Z_i$ can be calculated classically efficiently. This means various expectation values and thus different loss/cost functions can be estimated classically efficiently up to additive polynomial error. Therefore, the extended-IQP circuits can be classically trained using not just probability densities, but also a variety of cost functions based on measuring expectation values of different observables. This may be useful, for example when one had to sample bit strings which minimized a certain Hamiltonian.
%%%

%%%

%------ QCBM training -----------

\subsection{Training a QCBM efficiently classically}
In the QCBM setting, we estimate $p_{\mathrm{model}}(x)$ directly using classically-simulated output bit-strings, for fixed input $|0^{\otimes n}\rangle$. The amplitude to obtain a certain bitstring $x$ at the output of an extended-IQP circuit can be written as
\begin{equation}
    \Phi_x = \langle x |\hat{H}\hat{U}_2\hat{H}\hat{U}_1\hat{H}|0^{\otimes n}\rangle .
    \label{forr_qcbm}
\end{equation}
Writing $|x\rangle=(\prod_{i=1}^kX_i)|0^{\otimes n}\rangle$, where $i$ is indexed over locations where $1$ occurs in state $|x\rangle$. Thus we can write
\begin{equation}
\begin{split}
    \Phi_x &= \langle 0^{\otimes n}|(\prod_{i=1}^k\hat{X}_i)\hat{H}\hat{U}_2\hat{H}\hat{U}_1\hat{H}|0^{\otimes n}\rangle\\
    &=\langle 0^{\otimes n}|\hat{H}(\prod_{i=1}^k\hat{Z}_i)\hat{U}_2\hat{H}\hat{U}_1\hat{H}|0^{\otimes n}\rangle
\end{split}
    \label{forr_qcbm_2}
\end{equation}
\begin{figure*}[t]\centering
\includegraphics[scale=0.4]{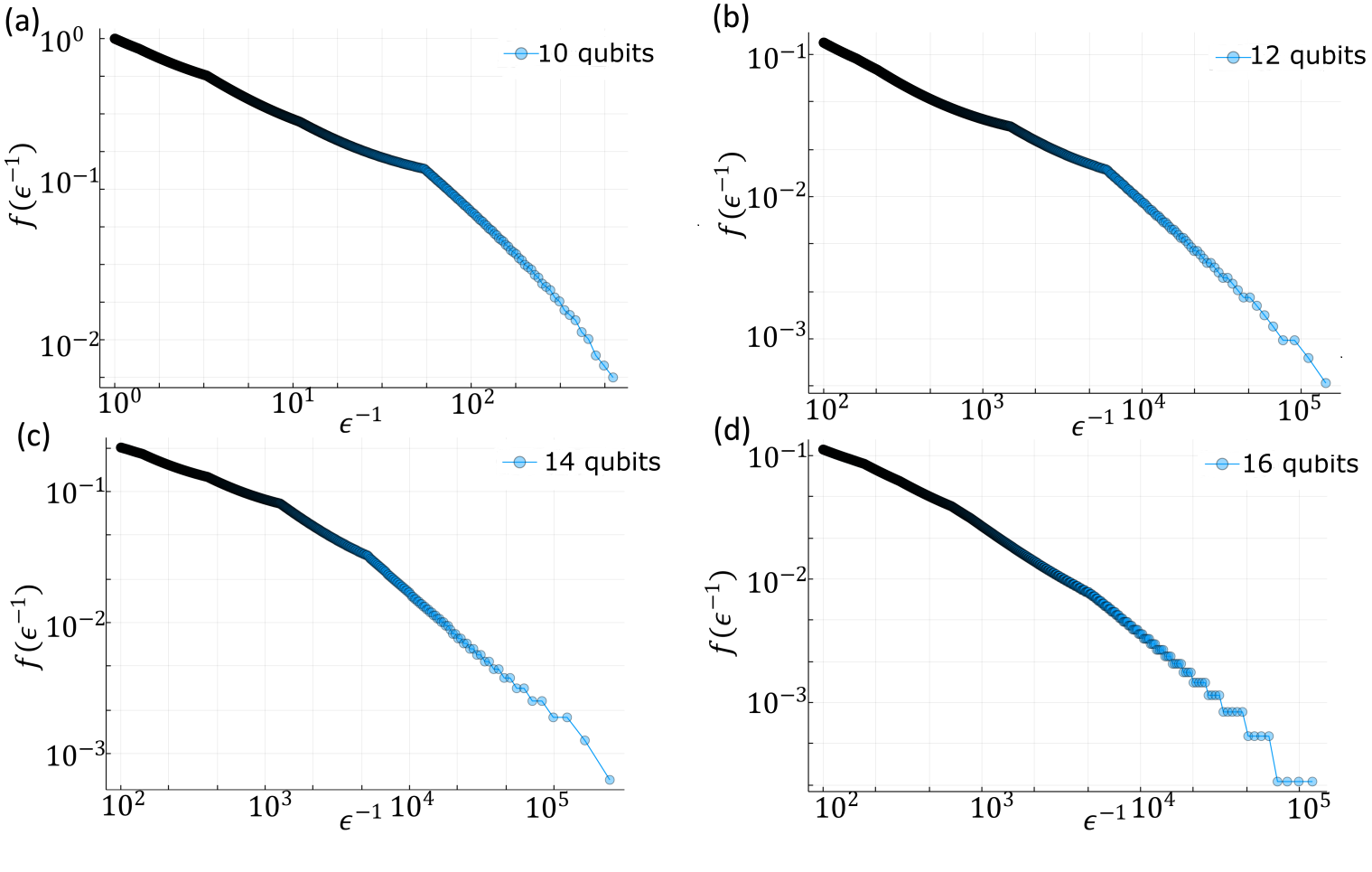}
\caption{Figure shows the plot of $f(\epsilon^{-1})$ vs $\epsilon^{-1}$ for a random probability distribution for different number of qubits (10,12,14 and 16) as a log-log plot. The downward curvature is an indication of super-polynomial behavior of $f(\epsilon^{-1})$.}
\label{t-sparse}
\end{figure*}
where we used the fact that $\hat{H}_i \hat{X}_i=\hat{Z}_i\hat{H}_i$. Absorbing the $\hat{Z}$ gates into $\hat{U}_2$, the above equation can be written as a Forrelation and thus can be computed classically efficiently. Using Eq. \ref{forr_qcbm_2}, $p_{\mathrm{model}}(x)$ can be computed. Similarly, just like Eq. \ref{Eq_dpdtheta}, the gradients with respect to $\theta$ can be written as
\begin{equation}
    \begin{split}
        \frac{\partial p_{\mathrm{model}}(x)}{\partial \theta_j}&=\mathrm{tr}\{|x\rangle\langle x|\hat{H}\hat{U}_{2,l:j+1}\hat{U}_{2,j}(\pi/2)\hat{\rho}_j\\
        &\hat{U}^\dagger _{2,j}(\pi/2)\hat{U}^\dagger_{2,l:j+1}\hat{H}]-\mathrm{tr}[|x\rangle \langle x|\hat{H}\\
        &\hat{U}_{2,l:j+1}\hat{U}_{2,j}(-\pi/2)\hat{\rho}_j
        \hat{U}^\dagger _{2,j}(-\pi/2)\hat{U}^\dagger_{2,l:j+1}\hat{H}\},\\
    \end{split}
    \label{Eq_dpdtheta_qcbm}
\end{equation}
where $\hat{\rho}_j=\hat{U}_{2,1:j}\hat{H}\hat{U}_1\hat{H}|0\rangle \langle 0|\hat{H}\hat{U}_1^\dagger \hat{H}\hat{U}_{2,1:j}^{\dagger}$. Since both the terms in the above equation are probabilities of obtaining a certain bitstring $x$ at the output of an extended-IQP circuit, using equations \ref{forr_qcbm},\ref{forr_qcbm_2}, they can be estimated classically. 
However, as described in the previous section, estimating gradients using parameter shift rule will require re-sampling from $P(y)$ for shifted parameters. Similar to the DQGM setting, we can estimate the gradients without the need for re-sampling for each parameters by replacing $\Phi$ with $\Phi_x$ and using Eqs.~\eqref{Eq_dforr2dtheta}-\eqref{Eq_diffMC}.

%-----  CIRCUIT and SAMPLING complexity

\subsection{Complexity of classical simulability: probabilities and sampling}

As discussed before, to enable classical training and hard sampling we need to check the properties of quantum circuits that we train. For this, we develop a workflow used for studying different properties of the chosen circuits (see the chart in Fig.~\ref{training_strategy}, right). We start selecting a family of circuits that allows additive polynomial estimation of probabilities. To show that it is still hard to sample from, we show that probabilities generated from these circuits are not poly-sparse \cite{Pashayan2020,VanDenNest2010}. We use two different approaches to show this. One approach involves numerical random sampling of these circuits and looking at their anti-concentration properties \cite{Hangleiter2018,Bermejo-Vega2018}. An output distribution of a unitary $\hat{U}$ for some setting of its parameters is said to anti-concentrate when
\begin{equation}
\mathrm{Pr}_{\hat{U}\sim\mu} \left(|\langle x|\hat{U}|0\rangle|^2 \ge \frac{\alpha}{N}\right) \ge \beta
\label{Eq_ac}
\end{equation}
for constants $\alpha,\beta$, where $N=2^n$ and $\hat{U}$ is drawn from a certain measure $\mu$. For example, \cite{Bermejo-Vega2018} shows a class of families for which $\beta=1/e$. The probability distributions of these circuits along with families discussed in Ref.~\cite{Boixo2018} converge to the Porter-Thomas distribution. In Ref.~\cite{Pashayan2020}, the authors prove that the anti-concentration and poly-sparsity cannot coexist. Thus, if we show that probability distributions from a family anti-concentrate, then we can conclude that the probability distributions are not poly-sparse and hence are hard to sample from.
\begin{figure*}[t]\centering
\includegraphics[scale=0.45]{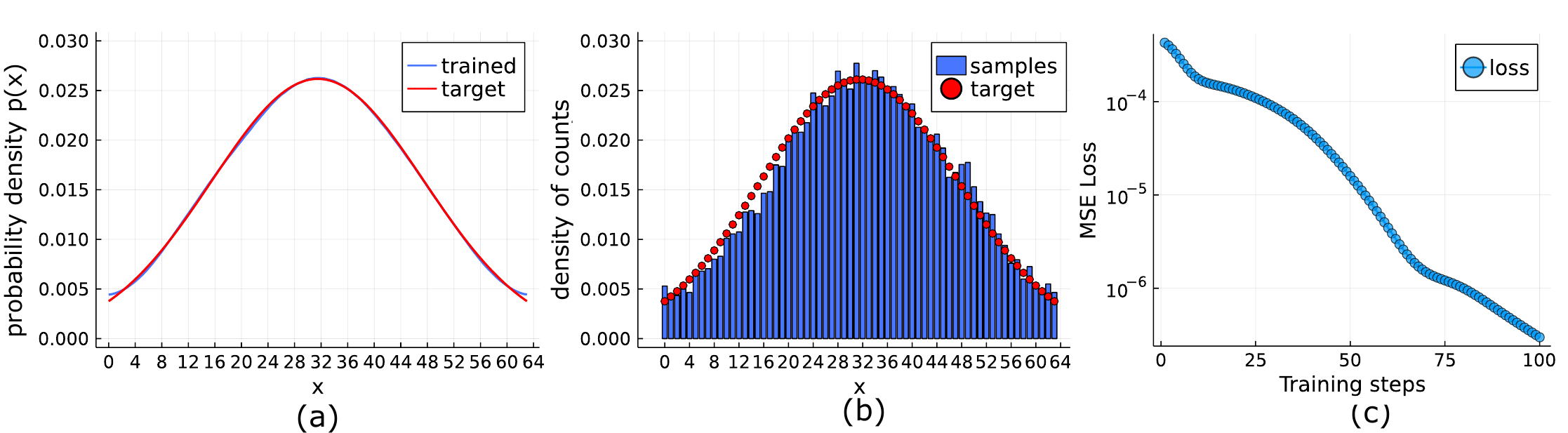}
\caption{Figure (a) shows the results for training QNN based on a extended-IQP circuit to generate a Gaussian probability density for 6 qubits. The figure shows an excellent fit between the trained and the target distribution. (b) shows the result of using the trained circuit in the training-stage to generate samples. The plot shows count density for 20,000 shots. (c) shows how the MSE loss goes down for 100 training steps.}
\label{training_result}
\end{figure*}
We use the approach discussed above for studying systems with up 20 qubits. For larger number of qubits we use the fact that the probability distributions converge to the Porter-Thomas distribution and use the cross-entropy difference to approximately measure the distance with the Porter-Thomas distribution. The cross-entropy difference is defined as
\begin{equation}
    \Delta H(p_{\mathrm{samp}}) \approx H_0 - \frac{1}{m}\sum_{j=1}^m\log\frac{1}{p_U({x_j^{\mathrm{samp}}})},
    \label{Eq_CE}
\end{equation}
where $H_0=\log(N) + \gamma$, and $\gamma\approx 0.577$ is Euler's constant. $p_U(x_j^{\mathrm{samp}})$ corresponds to the probability computed classically for the generated samples generated only. The error in $\Delta H(p_{\mathrm{samp}})$ is given by $\kappa/\sqrt{m}$, where $\kappa\approx m$. Thus, if we have a way of generating a finite number of samples, we can approximately characterize the distribution without the need of calculating all the probabilities. This is especially useful for larger registers ($n \ge 25$) where statevector calculations for all the probabilities (needed to measure anti-concentration or sparsity) rapidly becomes unfeasible. This approach has been used for classical benchmarking of data from random quantum circuits for $\sim$50 qubits \cite{Arute2019,Boixo2018}.   
%%%

To study resource requirements of various circuit families we use tensor networks to represent our quantum circuits, and we analyze their properties with classical simulation. Tensor networks use a tree-based decomposition to estimate the time-complexity of calculating probabilities. This is done by estimating the size of the largest tensor during the contraction process \cite{Markov2008}. The maximum size depends on the contraction order and various algorithms are used to find the contraction order which gives the smallest tensor size\cite{Pan2021,Kalachev2021,Gray2021}.

Apart from using anti-concentration, we can also measure whether a probability is poly-sparse or not. This is done by measuring the number of terms needed to $\epsilon$-approximate it with a sparse distribution. A $t$-sparse distribution, with only $t$ non-zero terms, can $\epsilon$-approximate a probability distribution $P(X)$ if and only if $\sum_x|P(x)-P_t(x)|\le \epsilon$ \cite{VandenNest2011}. Here, $P_t(x)$ is the probability distribution containing only the highest $t$ terms from $P_t(x)$ as the non-zero terms. We know that for $\epsilon=0$, $t=N$, where $N=2^n$ ($n$ is the number of qubits). Therefore, we can approximate the behavior of $t$ as $t(\epsilon)=N(1-f(1/\epsilon))$, where function $f$ shows a polynomial behavior if the distribution is poly-sparse. For an exponential behavior $f\sim e^{-1/\epsilon}$, the distribution is dense. Therefore, after calculating $t$ for different values of $\epsilon$, we calculate $f(1/\epsilon)=1-t/N$ and plot this as a function of $1/\epsilon$.

%=======   RESULTS ===============

\section{Results}

We proceed implementing the proposed strategies in practice. For enabling the classical training, we choose different quantum circuit families that include extended-IQP circuits compared with Product, Hadamard, IQP and IQP 1D-chain circuits (see corresponding diagrams in the Appendix). First, we compare the time-complexity for different families shown in Fig.~\ref{time_complexity}. These plots have been generated using Julia libraries \textsf{YaoToEinsum} for tensor network representation of quantum circuits built in \textsf{Yao}, which is based on the generic tensor contraction tool \textsf{OMEinsum} \cite{Luo2020,Kalachev2021,Gray2021,Kourtis2019}. As expected, for Product, Hadamard, 1D-chain the maximum size of the tensor during the contraction grows and quickly saturates (see inset in Fig.~\ref{time_complexity}). For IQP and extended-IQP, the time complexity grows linearly in the logarithmic scale. This implies that the classical computational complexity for calculating exact probabilities of the extended-IQP circuits, just like for IQPs, is exponential in the number of qubits.

Next, we study the anti-concentration properties of quantum circuits. Fig.~\ref{anticoncentration} shows the anti-concentration as a fraction of non-uniform probabilities compared to random circuits with bipartite connectivity. The randomness is chosen as follows. The first layer, the middle layer and the end layer are all composed of Hadamard gates. The $\hat{R}_{zz}(\theta)$ and the $\hat{R}_z(\theta)$ gates are chosen such that $\theta=k\pi/8$, with $k$ uniformly randomly chosen from $[0,1..,7]$ \cite{Bremner2016}. We observe that this set of gates approximates $\hat{U}$ drawn uniformly randomly from the Haar measure \cite{McClean2018BarrenLandscapes}. Specifically, we observe from Fig.~\ref{anticoncentration} that the fraction or probabilities $>1/2^n$ is very close to $1/e$ (dashed line), which is a good indicator that the probabilities do indeed anti-concentrate. The averaging is performed over 100 random circuits for each qubit number.

We can also show that probabilities for these circuits converge towards Porter-Thomas distribution for more number of qubits, with the variance reducing as well. Fig.~\ref{Porter_Thomas} shows the total variation distance measured for numerically obtained probability distributions for the extended-IQP circuit for 100 random configurations with respect to the Porter-Thomas distribution. The corresponding PDF is $P_{\mathrm{PT}}(p)=Ne^{-Np}$. Following Ref.~\cite{Bermejo-Vega2018}, the variation distance is defined as
\begin{equation}
    ||P-Q||_{\mathrm{TV}} :=\frac{1}{2}\sum_{X \in \Omega }|P(X)-1/m|,
\end{equation}
where we divide the set of probabilities into $m$ equally weighted bins $[p_0,...p_m]$ and take $\int_{p_i} ^ {p_{i+1}}P_{\mathrm{PT}}dp=1/m$. The set $\Omega$ is the set of probabilities in the interval $[p_i,p_{i+1}]$, where $i$ goes from 0 to $m$. $Q$ is the set of probabilities observed numerically over the set $\Omega$. We observe the distance rapidly approaching zero as the number of qubits is increased. 
\begin{figure*}[t]\centering
\includegraphics[scale=0.3]{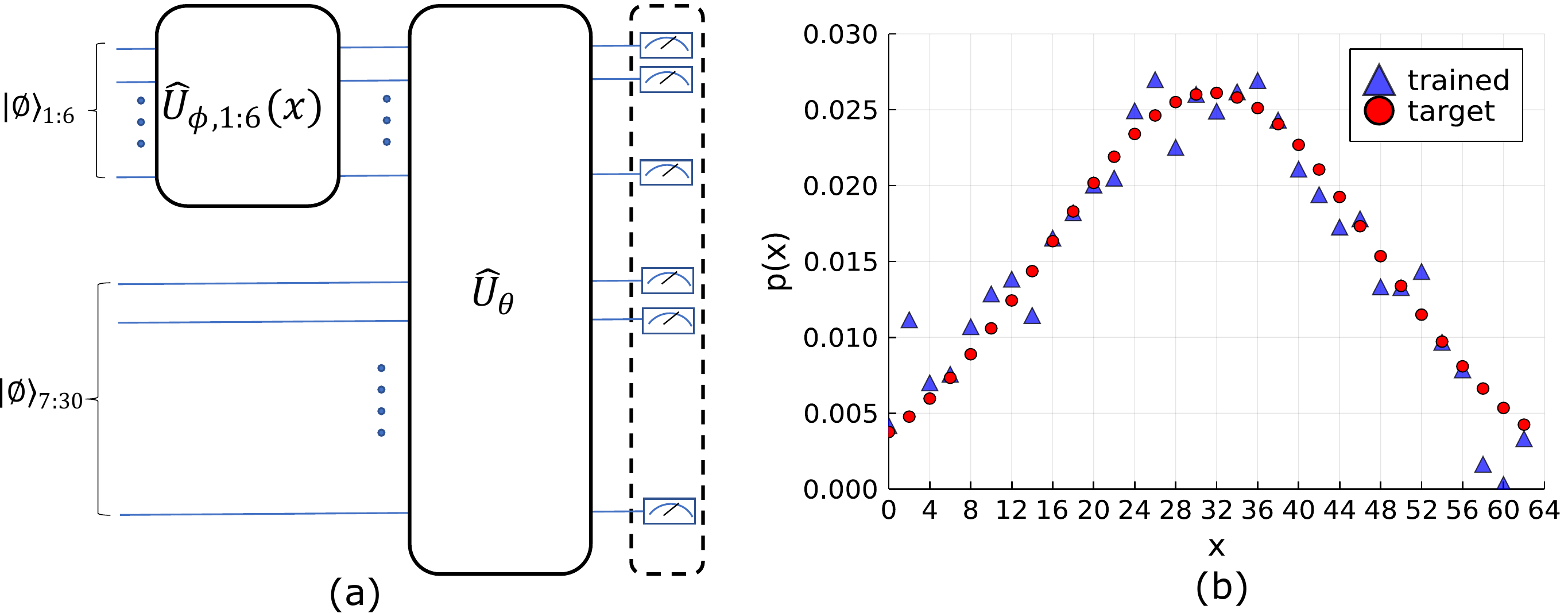}
\caption{(a) Setup used to train a probability distribution for 30 qubits. Only the parameters involving qubits 1-to-6 are updated during training while the remaining qubits are kept unmodified. (b) Results of training the DQGM circuit with an extended-IQP architecture to output a Gaussian probability distribution for $x$ between 1 to 64 with 32 equally spaced training points (integers label consecutive bitstrings). The estimates to the model probability density after training are obtained using Eq.~\eqref{Eq_px} with the obtained trained parameters. Loss values of ~$6.22\times 10^{-6}$ are reached at convergence.}
\label{training_30qubits}
\end{figure*}
We continue to study the sparsity. In Fig.~\ref{t-sparse}(a)-(d) we show the log-log plots of $f(1/\epsilon)$ vs $1/\epsilon$ for different number of qubits. The results are for a single random distribution. The downward curvature shows a super-polynomial decay rate, which indicates that the probability distribution is not poly-sparse.

Fig.~\ref{training_result} shows the results of training a quantum generative model as QNN based on the extended-IQP architecture for 6 qubits for a Gaussian probability density function. The circuit consists of initial phase feature map as a part of the extended-IQP architecture. Using the training stage as described in section II (Eq.~\ref{Eq_dqgm_train}), we try to maximize $P_{\mathrm{train}}(0^{\otimes n})$ for different values of $x$ by training $\hat{U}_{\theta}$. The cost function we use the mean square error, $\mathcal{L}=\sum_x |p_{\mathrm{model}}(x)-p_{\mathrm{target}}(x)|^2$. We see from Fig.~\ref{training_result}(a) that the trained QGM is able to closely follow the curve over the entire domain. The training has been performed using 128 points and a phase-feature map defined in Eq.~\eqref{Eq_fm}. We then use the trained circuit to generate samples using the sampling stage [Eq.~\eqref{Eq_dqgm_sampling}]. The results for 20,000 shots is shown in Fig.~\ref{training_result}(b).

We also performed training for $30$ qubits using the classical algorithms to estimate probabilities and their gradients based on Eqs.~\eqref{Eq_MC}-\eqref{Eq_diffMC}. To avoid issues related to barren plateau for training large number of qubits, we choose a particular probability distribution for training. Specifically we choose $p(x)$ to be a Gaussian probability density function from 0 to 63 and thus can be generated by effectively training the qubits 1 to 6. For qubits from 7 to 30, we apply the identity transformation as an initial setting. To do this, while still using the code for training 30 qubits, we apply the setup as shown in Fig.~\ref{training_30qubits}(a). The feature map is applied to qubits 1-to-6. To apply an identity transformation for the rest of the qubits we use the following fixed settings:\\
1. The $\theta s$ for all the 2-qubit gates involving qubits 7:30 are set to 0. \\
2. The $\theta s$ for all the single qubit gates are set to $\pi/2$ (which effectively sets the angle to $\pi/4)$. These gates along with the three $\hat{H}$ layers in the extended-IQP architecture, effectively implement the identity transformation on these qubits.

We now allow all the parameters to be trained (including for qubits 7-to-30). Starting with an initial identity transformation for qubits 7-30 ensures that the non-zero probability density largely remains confined to the events involving qubits 1-6 during the entire course of the training. Figure \ref{training_30qubits}(b) shows the results of training for 32 points and 100 training steps. The loss function value at convergence is ~$6.22\times 10^{-6}$. This simulation took approximately 42 hours on a regular desktop computer. While effectively the distribution is defined over 6 qubits, we stress that calculation of quantities like $R(y)$, sampling from $P(y)$ and calculation of $\Phi$ as defined in Eq.~\eqref{Eq_MC} and calculation of gradients involved all the 30 qubits.

\section{Conclusions and Outlook}
Our results show that certain circuit families, which we here call extended-IQP, can be trained classically by estimating probabilities up to an additive polynomial error, using the explicit generative modelling paradigm. We show that these circuits can be trained by estimating gradients classically in QCBM and DQGM settings. Using these techniques, we train a probability distribution for 30 qubits on a regular desktop computer. At the same time we show that these circuits still retain quantum advantage in terms of sampling. This we did by looking at the anti-concentration as well as the t-sparseness properties of the probability distributions up to 16 qubits. For higher number of qubits, cross-entropy benchmarking using samples based on tensor networks will be studied.

Recent work \cite{hinsche_learnability_2021} has highlighted the difficulty of training quantum generative models in the worst case, provided one has access only to estimates of quantities related to the target probability distribution. While in case of QCBM training using MMD loss this could be very important, in our case, however this issue does not arise since we are assuming knowledge of the target probability distribution.

So far we have focused on a single layer of Hadamards in the middle of commuting gates. But it may be possible to also extend these results to other single qubit operators. In addition it has been shown (Ref.~\cite{Bravyi2021}) that depth=2 QAOA circuits also allow additive polynomial estimation of the $\langle \psi|\hat{H}_{\mathrm{prob}}|\psi\rangle$, thus allowing classical training of these circuits while still showing quantum advantage in sampling. It could be an interesting possibility to classically simulate quantum annealing schedules to optimize annealing parameters. It would furthermore be interesting to study applications for this architecture for optimization problems \cite{Farhi2014, Varsamopoulos2022}.  

\begin{acknowledgements}

\emph{Acknowledgements.---}We thank QuiX Quantum for fruitful discussions.
\end{acknowledgements}

\emph{Disclosure.---}A patent application for the method described in this manuscript has been submitted by {\it{PASQAL}}.~\cite{patent}.

\bibliography{references}
\clearpage
\onecolumngrid
\appendix
\section{Circuits}
\begin{figure*}[hbt!]
    \centering 
    \includegraphics[scale=0.34]{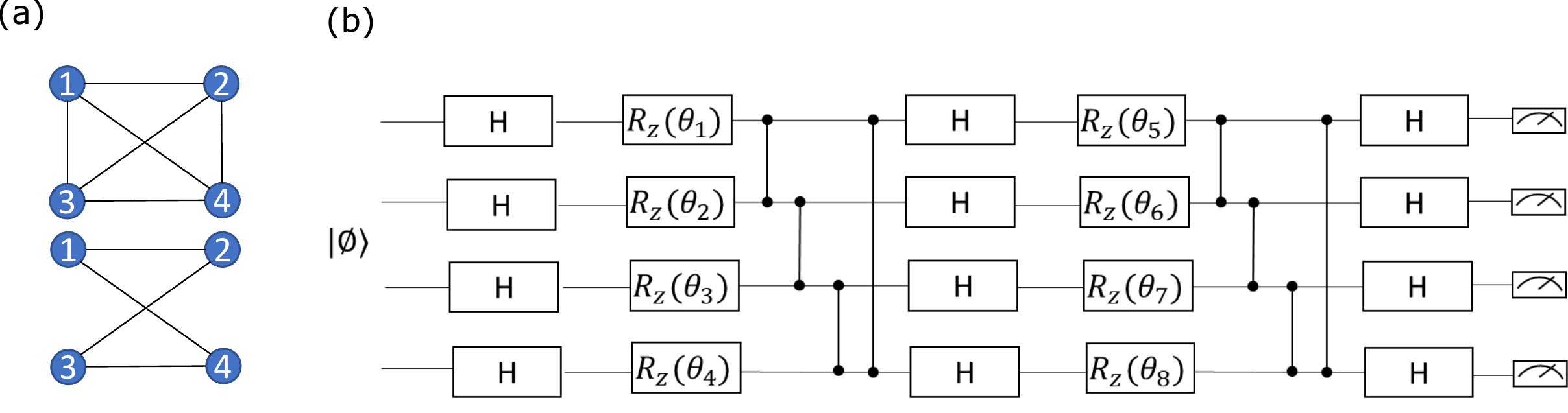}
    \caption{Figure (a) shows the qubit connectivity graph for an all to all and bipartite qubit connectivity. (b)shows the extended-IQP circuit for qubits with bipartite connectivity.\label{connectivity_graphs}}
\end{figure*}
\begin{figure*}[hbt!]
    \centering
    \includegraphics[scale=0.3]{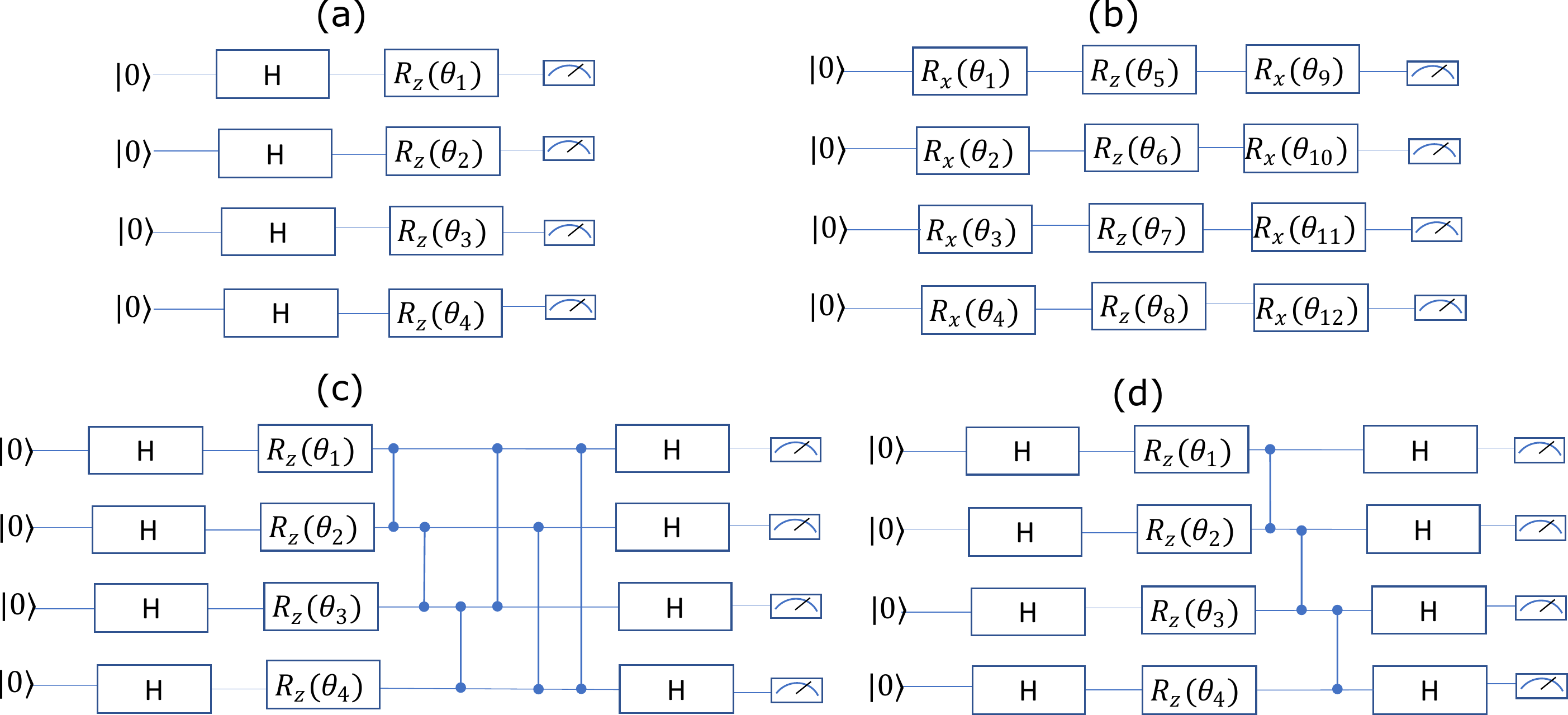}
    \caption{Figure shows different architectures studied (a)Hadamard (b)Product (c)IQP and (d)IQP 1D-chain.\label{circuit_families}}
\end{figure*}

\end{document}